\documentstyle[epsf,12pt]{article}

\begin{document}
\topmargin -35pt
\oddsidemargin 5mm

\newcommand {\beq}{\begin{eqnarray}}
\newcommand {\eeq}{\end{eqnarray}}
\newcommand {\non}{\nonumber\\}
\newcommand {\eq}[1]{\label {eq.#1}}
\newcommand {\defeq}{\stackrel{\rm def}{=}}
\newcommand {\gto}{\stackrel{g}{\to}}
\newcommand {\hto}{\stackrel{h}{\to}}
\newcommand {\1}[1]{\frac{1}{#1}}
\newcommand {\2}[1]{\frac{i}{#1}}
\newcommand {\th}{\theta}
\newcommand {\thb}{\bar{\theta}}
\newcommand {\ps}{\psi}
\newcommand {\psb}{\bar{\psi}}
\newcommand {\ph}{\varphi}
\newcommand {\phs}[1]{\varphi^{*#1}}
\newcommand {\sig}{\sigma}
\newcommand {\sigb}{\bar{\sigma}}
\newcommand {\Ph}{\Phi}
\newcommand {\Phd}{\Phi^{\dagger}}
\newcommand {\Sig}{\Sigma}
\newcommand {\Phm}{{\mit\Phi}}
\newcommand {\eps}{\varepsilon}
\newcommand {\del}{\partial}
\newcommand {\dagg}{^{\dagger}}
\newcommand {\pri}{^{\prime}}
\newcommand {\prip}{^{\prime\prime}}
\newcommand {\pripp}{^{\prime\prime\prime}}
\newcommand {\prippp}{^{\prime\prime\prime\prime}}
\newcommand {\pripppp}{^{\prime\prime\prime\prime\prime}}
\newcommand {\delb}{\bar{\partial}}
\newcommand {\zb}{\bar{z}}
\newcommand {\mub}{\bar{\mu}}
\newcommand {\nub}{\bar{\nu}}
\newcommand {\lam}{\lambda}
\newcommand {\lamb}{\bar{\lambda}}
\newcommand {\kap}{\kappa}
\newcommand {\kapb}{\bar{\kappa}}
\newcommand {\xib}{\bar{\xi}}
\newcommand {\ep}{\epsilon}
\newcommand {\epb}{\bar{\epsilon}}
\newcommand {\Ga}{\Gamma}
\newcommand {\rhob}{\bar{\rho}}
\newcommand {\etab}{\bar{\eta}}
\newcommand {\chib}{\bar{\chi}}
\newcommand {\tht}{\tilde{\th}}
\newcommand {\zbasis}[1]{\del/\del z^{#1}}
\newcommand {\zbbasis}[1]{\del/\del \bar{z}^{#1}}
\newcommand {\vecv}{\vec{v}^{\, \prime}}
\newcommand {\vecvd}{\vec{v}^{\, \prime \dagger}}
\newcommand {\vecvs}{\vec{v}^{\, \prime *}}
\newcommand {\alpht}{\tilde{\alpha}}
\newcommand {\xipd}{\xi^{\prime\dagger}}
\newcommand {\pris}{^{\prime *}}
\newcommand {\prid}{^{\prime \dagger}}
\newcommand {\Jto}{\stackrel{J}{\to}}
\newcommand {\vprid}{v^{\prime 2}}
\newcommand {\vpriq}{v^{\prime 4}}
\newcommand {\vt}{\tilde{v}}
\newcommand {\vecvt}{\vec{\tilde{v}}}
\newcommand {\vecpht}{\vec{\tilde{\phi}}}
\newcommand {\pht}{\tilde{\phi}}
\newcommand {\goto}{\stackrel{g_0}{\to}}
\newcommand {\tr}{{\rm tr}\,}
\newcommand {\GC}{G^{\bf C}}
\newcommand {\HC}{H^{\bf C}}
\newcommand{\vs}[1]{\vspace{#1 mm}}
\newcommand{\hs}[1]{\hspace{#1 mm}}

\setcounter{page}{0}

\begin{titlepage}

\begin{flushright}
TIT/HEP-454\\
OU-HET 359\\
hep-th/0008240\\
August 2000
\end{flushright}
\bigskip

\begin{center}
{\LARGE\bf Auxiliary Field Formulation of 
Supersymmetric Nonlinear Sigma Models
}\footnotetext[1]{
Talk given by M. N. at 
XXXth International Conference on High Energy Physics  
(ICHEP 2000), July 27-August 2, 2000, Osaka, Japan
}
\vs{10}

\bigskip
{\renewcommand{\thefootnote}{\fnsymbol{footnote}}
{\large\bf Kiyoshi Higashijima$^a$\footnote{
     E-mail: higashij@phys.sci.osaka-u.ac.jp.}
 and Muneto Nitta$^b$\footnote{
E-mail: nitta@th.phys.titech.ac.jp}
}}

\setcounter{footnote}{0}
\bigskip

{\small \it
$^a$Department of Physics,
Graduate School of Science, Osaka University,\\
Toyonaka, Osaka 560-0043, Japan\\
$^b$Department of Physics, Tokyo Institute of Technology, 
Oh-okayama, \\ Meguro, Tokyo 152-8551, Japan\\
}

\end{center}
\bigskip

\begin{abstract}
Two dimensional ${\cal N}=2$ 
supersymmetric nonlinear sigma models on
hermitian symmetric spaces 
are formulated in terms of the auxiliary superfields.
If we eliminate auxiliary vector and chiral superfields,  
they give D- and F-term constraints to define the target manifolds. 
The integration over auxiliary vector superfields, which can be 
performed exactly, is equivalent to the elimination of the auxiliary 
fields by the use of the classical equations of motion.
\end{abstract}

\end{titlepage}


\section{Introduction}
Two dimensional nonlinear sigma models 
(NL$\sigma$M) have been 
interested in, since they have many similarities to 
four dimensional QCD such as 
asymptotic freedom, the mass gap, instantons and so on. 
Non-perturbative analyses of 
NL$\sigma$M can be easily done 
by the large-$N$ method. 
In the $O(N)$ model,  
the mass gap appears as non-perturbative effect. 
In the ${\bf C}P^{N-1}$ model, 
a gauge boson is dynamically generated. 
${\cal N}=1$ SUSY NL$\sigma$M (SNL$\sigma$M) 
have been also investigated. 
The ${\cal N}=1$ $O(N)$ model is simply a combination of 
the bosonic $O(N)$ NL$\sigma$M and the Gross-Neveu model 
which shows dynamical chiral symmetry breaking~\cite{WiAl}.  

Along this line it is interesting to discuss 
non-perturbative analyses of ${\cal N}=2$ SNL$\sigma$M, 
since they may have similarities 
to four dimensional ${\cal N}=1$ QCD. 
To discuss non-perturbative effects 
of NL$\sig$M by the large-$N$ method, 
it is necessary to reformulate them 
by the auxiliary field method. 
However there was no auxiliary field formulation 
of ${\cal N}=2$ SNL$\sigma$M 
except for the ${\bf C}P^{N-1}$ and 
the Grassmann models~\cite{WiDDL,Ao}. 
In this talk, we formulate ${\cal N}=2$ SNL$\sig$M on 
hermitian symmetric spaces (HSS) $G/H$ (see Table~\ref{HSS}) 
by the auxiliary field method~\cite{HN1,HN2,HN3}. 
Since ${\cal N}=2$ SUSY in two dimensions 
is equivalent to ${\cal N}=1$ SUSY in four dimensions, 
we use the notation of four dimensions. 
\begin{table*}[h]
\caption{Hermitian symmetric spaces (HSS). 
\label{HSS}
}
\begin{center}
\begin{tabular}{|c|c|c|}
 \noalign{\hrule height0.8pt}
  Type & $G/H$ & $\dim_{\bf C} (G/H)$\\
 \hline
 \noalign{\hrule height0.2pt}
 AIII$_1$&${\bf C}P^{N-1}=SU(N)/SU(N-1)\times U(1)$&$N-1$\\
 AIII$_2$&$G_{N,M}({\bf C})=U(N)/U(N-M)\times U(M)$  &$M(N-M)$\\
 BDI     &$Q^{N-2}({\bf C})=SO(N)/SO(N-2)\times U(1)$&$N-2$\\
 CI      &$Sp(N)/U(N)$                          &$\1{2}N(N+1)$\\
 DIII    &$SO(2N)/U(N)$                         &$\1{2}N(N-1)$\\     
 EIII    &$E_6/SO(10)\times U(1)$                    &$16$\\
 EVII    &$E_7/E_6 \times U(1)$                      &$27$\\   
 \noalign{\hrule height0.8pt}
 \end{tabular}
 \end{center}
\end{table*}

\section{Auxiliary Field Formulation}
\subsection{SNL$\sigma$M 
without F-term constraints}
It was recognized in Ref.~\cite{WiDDL} 
that the ${\bf C}P^{N-1}$ model can be 
constructed by an auxiliary vector superfield $V$:
Let $\phi$ be dynamical chiral superfields belonging to 
${\bf N}$ of $SU(N)$. 
Then its K\"{a}hler potential can be written as 
\beq
 K(\phi,\phi^{\dagger},V) 
 = e^V \phi^{\dagger} \phi -cV, \label{gauged-Kahler}
\eeq
where $c V$ is an Fayet-Iliopoulos (FI) D-term. 
($c$ is a positive constant called an FI-parameter.)  

This model can be immediately 
generalized to the Grassmann model, 
$G_{N,M}({\bf C})$ $(N>M)$ 
by replacing $\phi$ by an $N \times M$ 
matrix chiral superfield $\Phi$ 
and $V$ by an $M \times M$ matrix vector superfield 
$V = V^A T_A$, where $T_A$ are generators of 
$U(M)$ gauge group:~\cite{Ao}    
\beq
 K (\Ph,\Ph\dagg,V) = \tr(\Ph\dagg\Ph e^V) - c\, \tr V.  
     \label{gauged-Kahler2}
\eeq

After we integrate out $V$ and fix a gauge 
in these models,  
we obtain K\"{a}hler potentials 
of the Fubini-Study metric 
of ${\bf C}P^{N-1}$ and its generalization to 
the Grassmann manifold.

\subsection{SNL$\sigma$M 
with F-term constraints}
In this section, to obtain the rest of HSS,  
we introduce auxiliary chiral superfields 
$\phi_0$ or $\Phi_0$ besides auxiliary vector superfields 
and construct $G$-invariant superpotentials 
as summarized in Table~\ref{F-con.}~\cite{HN1}.   
\begin{table*}[t]
\caption{F-term constraints and embedding.\label{F-con.}}
\begin{tabular}{|c|c|c|c|c|}
 \noalign{\hrule height0.8pt}
  $G/H$ & $G$-invariants & superpotentials & constraints 
  & embedding \\
 \hline
 \noalign{\hrule height0.2pt}
  ${SO(N) \over SO(N-2)\times U(1)}$ 
      & $I_2 = \phi^2$     
      & $\phi_0 I_2$ & $I_2=0$ & ${\bf C}P^{N-1}$\\
  ${SO(2N)\over U(N)}$, ${Sp(N)\over U(N)}$ 
      & ${I_2}\pri =\Phi^T J \Phi$ 
      & $\tr (\Phi_0 {I_2}\pri)$ 
      & ${I_2}\pri = 0$ & $G_{2N,N}$\\
  ${E_6\over SO(10)\times U(1)}$ 
      & $I_3=\Gamma_{ijk}\phi^i\phi^j\phi^k$ 
      & $\Gamma_{ijk}{\phi_0}^i\phi^j\phi^k$ 
      & $\del I_3 =0$ & ${\bf C}P^{26}$\\
  ${E_7\over E_6 \times U(1)}$ 
      & $I_4=d_{\alpha\beta\gamma\delta}
         \phi^{\alpha}\phi^{\beta}
         \phi^{\gamma}\phi^{\delta}$ 
      & $d_{\alpha\beta\gamma\delta}
         {\phi_0}^{\alpha}\phi^{\beta}
         \phi^{\gamma}\phi^{\delta}$  
      & $\del I_4 =0$ & ${\bf C}P^{55}$\\  
 \noalign{\hrule height0.8pt}
 \end{tabular}
\end{table*}
Integration over the auxiliary chiral superfields 
gives F-term constraints, which are holomorphic. 

The simplest example is 
$Q^{N-2}({\bf C})$, 
where dynamical fields constitute an $SO(N)$ vector $\phi$. 
It can be embedded into ${\bf C}P^{N-1}$ by 
an F-term constraint $\phi^2=0$. 
Hence an auxiliary chiral superfield is taken to be 
a singlet $\phi_0$ and a superpotential to be  
$W = \phi_0 \phi^2$. 

Both $SO(2N)/U(N)$ and $Sp(N)/U(N)$ can be embedded into 
$G_{2N,N}$ by F-term constraints $\Phi^T J \Phi = 0$, 
where $\Phi$ is a $2N \times N$ matrix chiral superfield.   
Here $J= \pmatrix {{\bf 0}&{\bf 1}_N \cr
       \epsilon{\bf 1}_N &{\bf 0} }$,  
where $\epsilon = +1$ (or $-1$) for $SO(2N)$ (or $Sp(N)$).
Then auxiliary chiral superfields constitute 
an $N \times N$ matrix $\Phi_0$ belonging 
to the symmetric (anti-symmetric) tensor representation of 
$SO(2N)$ ($Sp(N)$), and the superpotential is 
$W = \tr(\Phi_0 \Phi^T J \Phi)$. 

It is less trivial to find F-term constraints for 
$E_6$ and $E_7$ models. 
The F-term constraints are $G$-invariants   
($I_2$ or ${I_2}\pri$ in Table \ref{F-con.})
for the classical groups; on the other hand 
they are not $G$-invariants but 
the differentiation of $G$-invariants
($\del I_3$ or $\del I_4$ in Table \ref{F-con.}) 
for the exceptional groups.  
We must introduce auxiliary chiral superfields 
${\phi_0}^i$ and ${\phi_0}^{\alpha}$ 
($i=1,\cdots,27$; ${\alpha}=1,\cdots,56$) 
belonging to the fundamental representations  
of $E_6$ and $E_7$. 
Superpotentials can be written as 
$W= \Gamma_{ijk} {\phi_0}^i \phi^j \phi^k$ 
and $W= d_{\alpha\beta\gamma\delta} 
{\phi_0}^{\alpha}\phi^{\beta}\phi^{\gamma}\phi^{\delta}$ 
for $E_6$ and $E_7$ models,  
where $\Gamma$ and $d$ are
rank-$3$ and rank-$4$ symmetric tensors of 
$E_6$ and $E_7$, respectively.  
At first sight one might consider 
the number of the F-term constraints are too large. 
However some of them are not independent due to 
the identities of invariant tensors $\Gamma$ and $d$. 
(Only $10$ of $27$ equations and $28$ of $56$ equations 
are independent for $E_6$ and $E_7$ cases, respectively.)

\section{Integration over Auxiliary Fields}
The path integration over auxiliary chiral superfields 
$\phi_0$ or $\Phi_0$ is 
easy since they are linear in the lagrangian; 
on the other hand, 
the path integration over auxiliary vector superfields $V$ 
is nontrivial. 
However we can perform integration over $V$ {\it exactly} 
although they are not quadratic~\cite{HN2}.  
The result for Abelian $V$ is  
\beq
  \int [d V] \exp \left[i \int d^D x d^4 \theta  
      \left( \phi\dagg\phi \,e^V 
            - c\, V \right) \right]  
 = \exp \left[ i \int d^D x d^4 \theta
                   \,c \,\log (\phi\dagg\phi) \right] . 
    \label{IOV}
\eeq
This coincides with the result obtained by the 
equation of motion of $V$. 
Its coincidence is highly nontrivial 
since there are infinite number of corrections 
in bosonic cases. 
Eq.~(\ref{IOV}) can be proved 
by the following theorem.\\  
{\bf Theorem.} 
Let $\sig (x,\theta,\bar{\theta})$ and 
$\Phi (x,\theta,\bar{\theta})$ 
be vector superfields and $W$ be a function of $\sig$. 
Then,  
\beq
 \int [d\sig] \exp\left[i \int d^Dx d^4\theta  
    \;(\sig \Phi - W(\sig)) \right] 
 = \exp\left[i \int d^Dx d^4\theta  \;U(\Phi) \right] .\;\;\;
 \label{Quantum_Legendre}
\eeq
Here $U(\Phi)$ is defined as 
$U(\Phi) = \hat \sig(\Phi) \Phi 
  - W(\hat\sig(\Phi))$, 
where $\hat{\sig}$ is a 
solution of the stationary equation 
${\del \over \del \sig} 
  (\sig \Phi - W (\sig)) |_{\sig=\hat \sig} 
 = \Phi - W\pri (\hat \sig) = 0$.\\  
To prove Eq.~(\ref{IOV}), 
put $\Phi = \phi\dagg\phi$, $\sig = e^V$ and $W= c\log \sig$, 
and note that $[d\sig] = [dV]$. 
This theorem can be generalized to 
many variables and to matrix variables,  
which can be applied to 
integration over non-Abelian vector superfields.  
As an application of the theorem, we can show that 
\beq
 && \int [d V] \exp \left[i \int d^D x d^4 \theta  
      \left( f (\phi\dagg\phi \,e^V) 
            - c\, V \right) \right] \non
 && =\int [d V] \exp \left[i \int d^D x d^4 \theta  
      \left( \phi\dagg\phi \,e^V 
            - c\, V \right) \right],
\eeq
where $f$ is an {\it arbitrary} function.

\section{Discussion}
Non-perturbative analyses 
of ${\cal N}=2$ SNL$\sig$M 
on HSS are possible, 
which are in progress~\cite{HKNT}. 

Let us discuss a generalization to 
an arbitrary K\"{a}hler $G/H$. 
It is known that $H$ must be of the form 
$H = H_{\rm s.s} \times U(1)^n$, 
where $H_{\rm s.s}$ is the semi-simple subgroup of $H$ and 
$n = {\rm rank}\, G - {\rm rank}\, H_{\rm s.s}$, 
and a K\"{a}hler potential 
has $n$ free parameters~\cite{SNLR}.   
As seen in Eq.~(\ref{IOV}), 
the FI-parameter $c$ represents a size of 
$G/H$ after integration over $V$.  
Hence to obtain a K\"{a}hler $G/H$, 
it is needed to introduce $n$ FI-terms by considering 
a gauge group including $n$ Abelian factors. 

\section*{Acknowledgements}
The work of M.~N. is supported in part 
by JSPS Research Fellowships.


\end{document}